# "Openness of Search Engine": A Critical Flaw in Search Systems; A Case study on Google, Yahoo and Bing

Katuru SM Kalyana Chakravarthy

ISTQB Certified QA Engineer

Hyderabad, A.P, India

Kalyan.Katuru@gmail.com

*ABSTRACT*

There is no doubt that Search Engines are playing a great role in Internet usage. But all the top search engines Google, Yahoo and Bing are having a critical flaw called "Openness of a Search Engine". An Internet user should be allowed to get the search results only when requested through Search engine's web page but the user must not be allowed to get the search results when requested through any web page that does not belong to the Search Engine. Only results of a search engine should be available to the Internet user but not the Search Engine. This paper explains the critical flaw called "Openness of Search Engine" with a case study on top 3 search engines 'Google', 'Yahoo' and 'Bing'. This paper conducts an attack based test using J2EE framework and proves that 'Google' passed the test and it strongly protects its Critical Search System, where 'Yahoo' and 'Bing' are failed to protect their Search Engines. But previously 'Google' also had other high severity issues with the Openness of search engine; this paper reveals those issues also. Finally this paper appeals strongly to the all top Search Engines to fix their critical flaws of "Openness of Search Engine".

## 1. INTRODUCTION

### 1.1 What is Openness of a Search Engine?

This paper defines Openness of a Search Engine as "Ability of a search Engine to provide the Search results to any HTTP page's request".

### 1.2 Critical loss with openness of a Search Engine

A flaw with openness of a Search Engine is that user can retrieve the search results by requesting from any HTML page that does not belong to search engine. When user is able to retrieve the search results by requesting from any HTML page, he can modify the branding simply and can own the search results with his branding. Critical loss with Openness of Search Engine is that hacker's engine can display the search results of an original search engine with fake branding. This causes hacker to own a complete search engine by just developing own branded HTML page and simple java code.

## 1.3 What is wrong?

"Search Engine should not be open to a HTML page that does not belong to the Search Engine".

## 1.4 Technique to hack the Search Results

In this paper a technique called "Own a Search Engine" is proposed to prove the Openness of top 3 Search Engines "Google, Yahoo and Bing". Following figure gives the overview of the proposed technique to hack the search results from a search engine and to display with the fake branding.

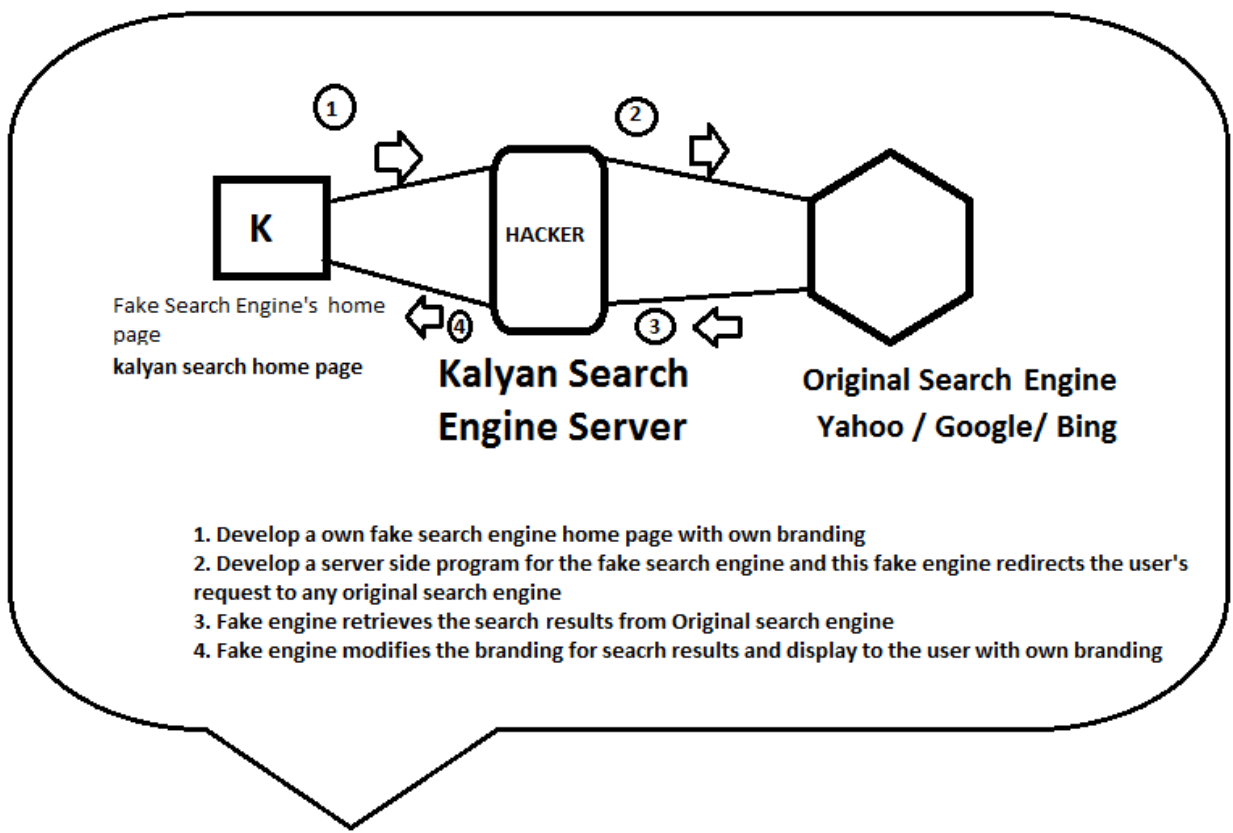

Fig 1.1 Overview of proposed technique to hack search results

## 1.5 How to solve this critical issue

Google has the solution but others do not have. But Google also had some loop holes. Following sections discuss all these issues in a detailed manner. Finally this paper strongly appeals to all top search engines to fix this critical issue.

# 2. OWN A SEARCH ENGINE

## 2.1 A proposed technique "Own A Search Engine" to hack the search results of a search engine

This paper proposes a technique called "Own A Search Engine" to hack the search results of a Search Engine and to display with own branding.

### 2.1.1 Steps involved in the technique "Own a Search Engine"

**Setp-1**:

Develop a search engine client HTML Page called 'ownsearchengine.html'.

Following is the designed page:

<HTML>

<HEAD>

</HEAD>

<BODY bgcolor=pink>

<CENTER>

<form action="./ownsearchengine.jsp">

<H1>KALYAN SEARCH ENGINE</H1>

<Input type=text name=searchstring>

<Input type=submit>

</CENTER>

</BODY>

</HTML>

**Setp-2**: Develop a server side program (a JSP) named as 'ownsearchengine.jsp' and write the pieces of code in the JSP such that it should do following actions.

**Step-2a:** Read the input string from the page 'ownsearchengine.html'.

Following is the piece of code that can be used to read the input string from the page 'ownsearchengine.html'

    String ss=request.getParameter("searchstring");

Here "searchstring" is the Input parameter name that is passed to the JSP from the HTML page.

**Step-2b:** Redirect the input string to an original search engine.

Following is the piece of code that can be used to redirect the input string to an original search engine.

```
String url="http://search.yahoo.com/search?q="+ss;

URL urlobj= new URL(url);

URLConnection uc =urlobj.openConnection();
```

This step is heart of this technique. Generally all search engines' request uses 'q' as input name for search string. So if the search engine is called by providing the "q" as parameter in the url, search results can be retrieved.

**Step-2c:** Catch the search results from the original search engine.

Following is the piece of code that can be used to catch the search results from the original search engine

```
BufferedReader in = new BufferedReader(

new InputStreamReader(

uc.getInputStream()));

String inputLine="", output= "";

while ((inputLine = in.readLine()) != null)

output=output+inputLine;

in.close();
```

**Step-2d:** Modify the branding of the search results to display own branding. Following is the piece of code that can be used to modify the branding of the search results to display own branding

```
output=output.replaceAll("Yahoo!", "Kalyan!")

/* This is example code to replace branding of yahoo search results*/
```

**Step-2e**: Display the hacked search results to the end user with own branding. Following is the piece of code used to display the hacked search results to the end user with own branding

```
<%=output %>
```

## 3. CASE STUDY-I: OWN THE ENGINE "SEARCH.YAHOO.COM"

In this section proposed technique "Own a Search Engine" is applied to hack the search results of Seach.Yahoo.com. First observe the result of Yahoo search engine when searched for some test string "sample".

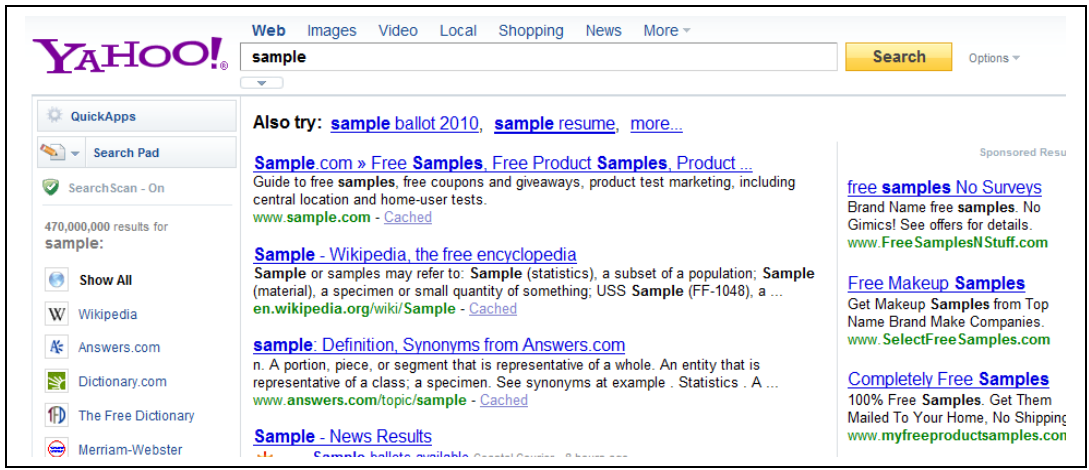

Fig 3.1 Original results of Yahoo search engine

Now apply the "Own a Search Engine" technique, when user access a fake search engine and tried to search for a string "sample", the fake engine internally redirects the search request to the original Yahoo engine and retrieve the results and modify the branding of results to display own branding. Finally user will see the results of Yahoo search engine with the fake branding.

Following figures explains the hacking of Yahoo search results.

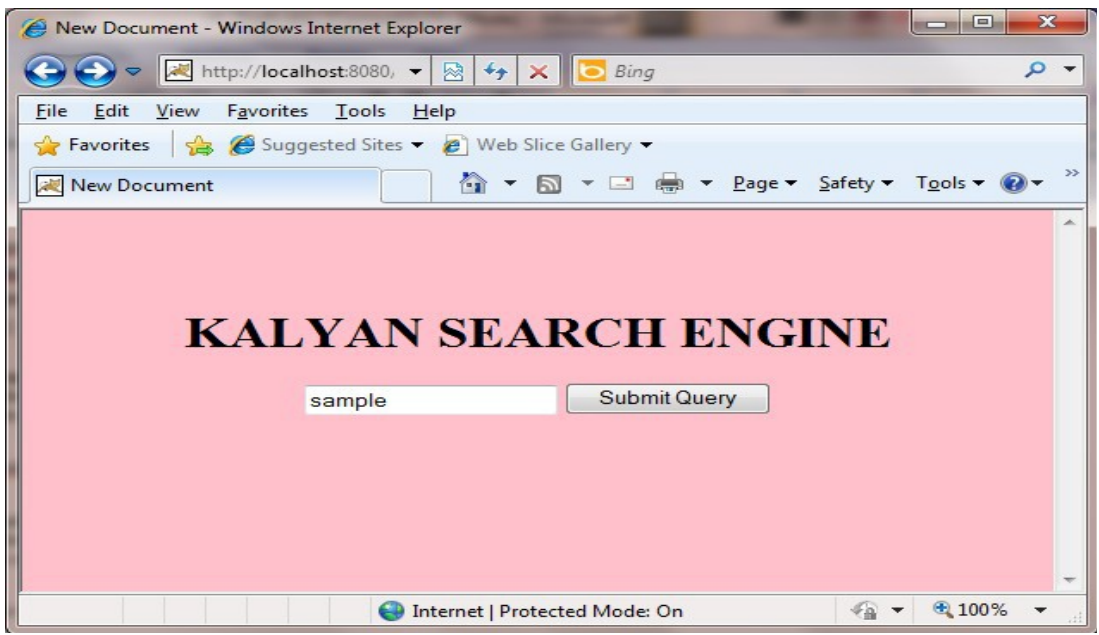

Fig 3.2 User tries to search from a fake engine

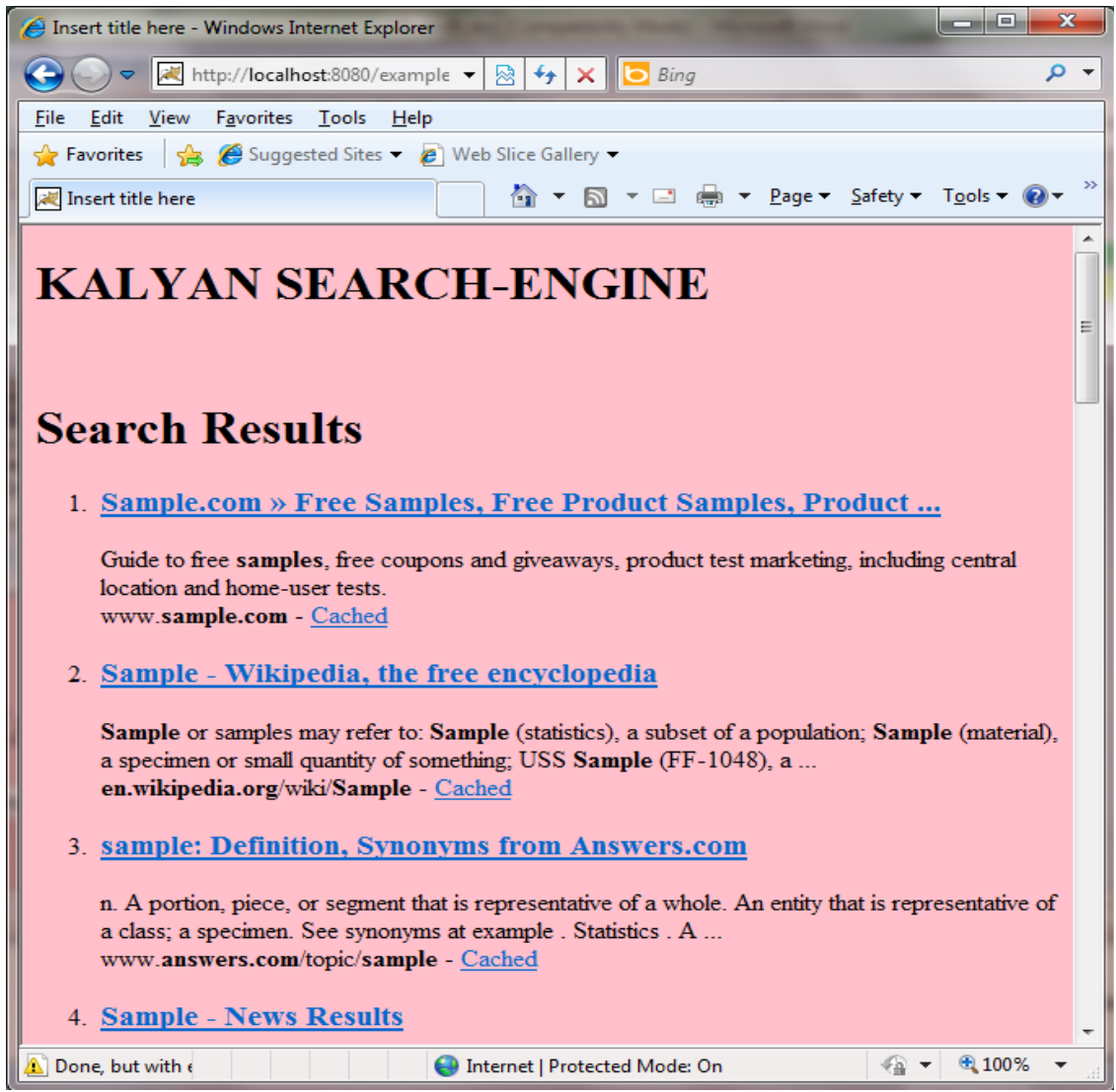

Fig 3.3 Fake engine internally uses Yahoo search engine and display results with fake branding

Compare the search results displayed in fig 3.1and 3.3. It is observed that same results are shown in the both cases because fake engine internally uses Yahoo search engine. From the above figures it is proved that Yahoo search engine is open to any HTML page and any user can hack the results of Yahoo and can show the Yahoo results with fake branding.

Hence this paper is concluding that:

"Yahoo failed to protect its Search Engine from Openness".

Yahoo search engine suffers from the Openness of the Search Engine. This paper strongly appeals to Yahoo to fix this issue, else any user can hack Yahoo's results. I.e. indirectly hacker owns a search engine which is Yahoo in real.

## 4. CASE STUDY-II: OWN THE ENGINE "BING.COM"

In this section proposed technique "Own a Search Engine" is applied to hack the search results of Bing.com. First observe the result of Bing search engine when searched for some test string "sample".

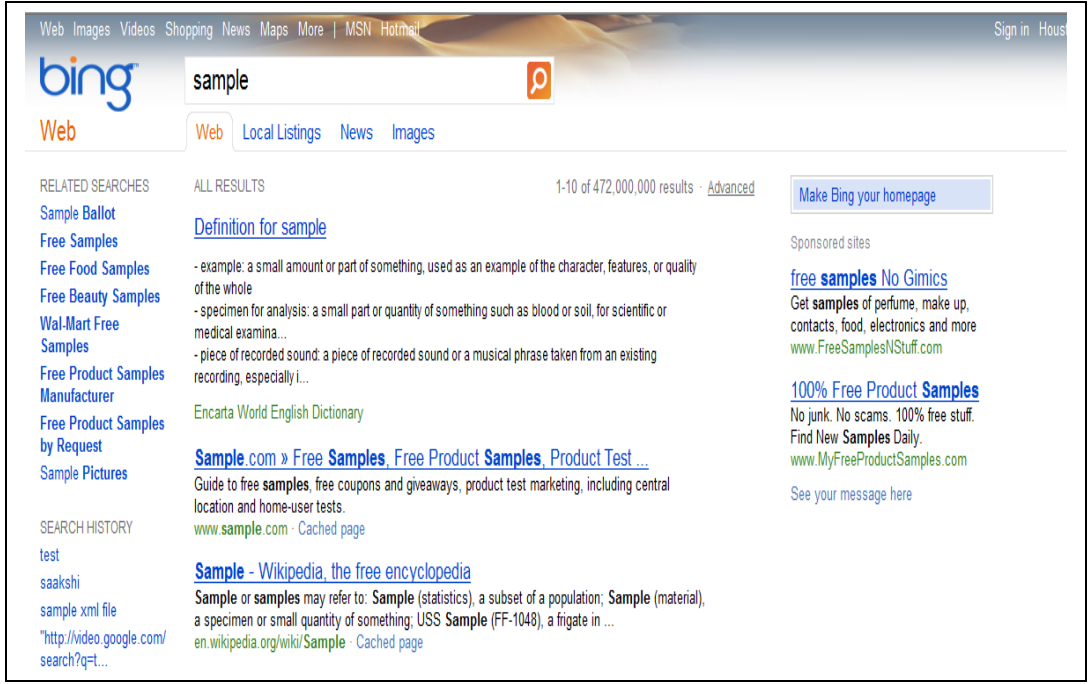

Fig 4.1 Original results of Bing search engine

Now apply the "Own a Search Engine" technique, when user access a fake search engine and tried to search for a string "sample", the fake engine internally redirects the search request to the original Bing engine and retrieve the results and modify the branding of results to display own branding. Finally user will see the results of Bing search engine with the fake branding.

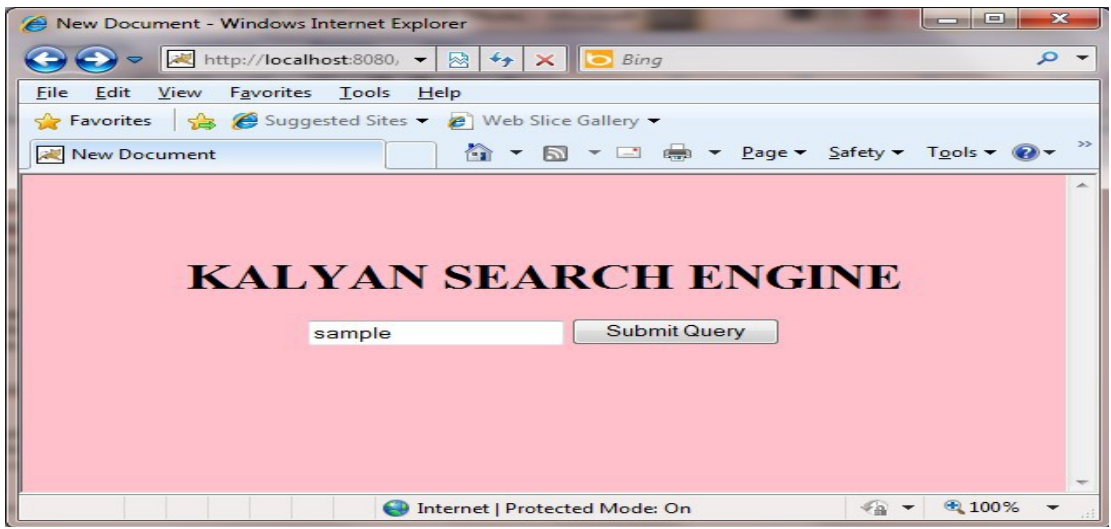

Fig 4.2 User tries to search from a fake engine

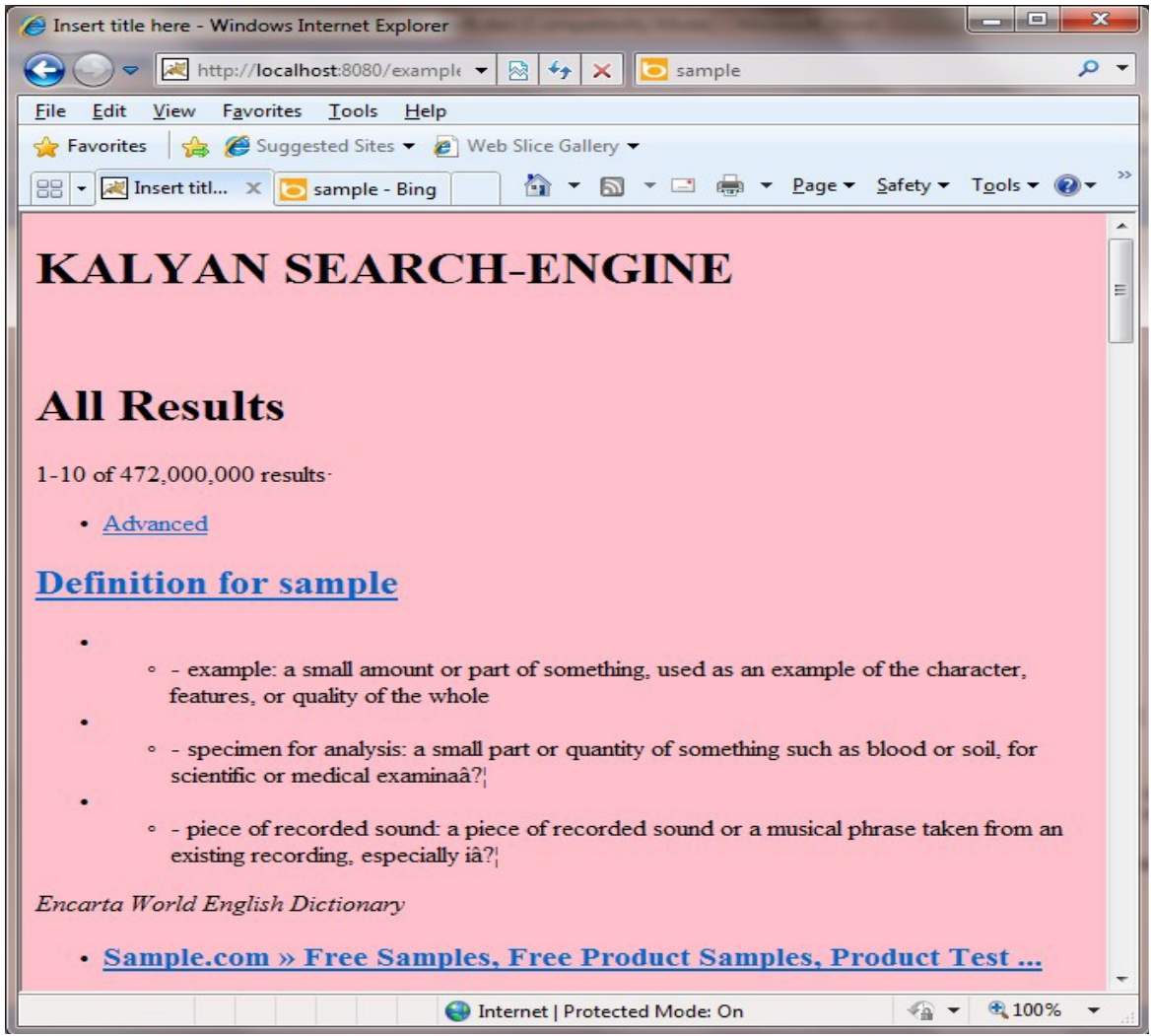

Fig 4.3 Fake engine internally uses Bing search engine and displays the results with fake branding.

Compare the search results displayed in fig 4.1and 4.3. It is observed that same results are shown in the both cases because fake engine internally uses Bing search engine.

From the above figures it is proved that Bing search engine is open to any HTML page and any user can hack the results of Bing and can show the Bing results with fake branding.

Hence this paper is concluding that:

"Bing failed to protect its Search Engine".

## 5. CASE STUDY-III: OWN THE ENGINE "NEWS.GOOGLE.COM"

Though Google is strongly protecting its search engine, it had some loop holes in the past. Google was failed to protect its other search applications like "News.Google.com".

First observe the result of news.google.com when searched for some test string "sample".

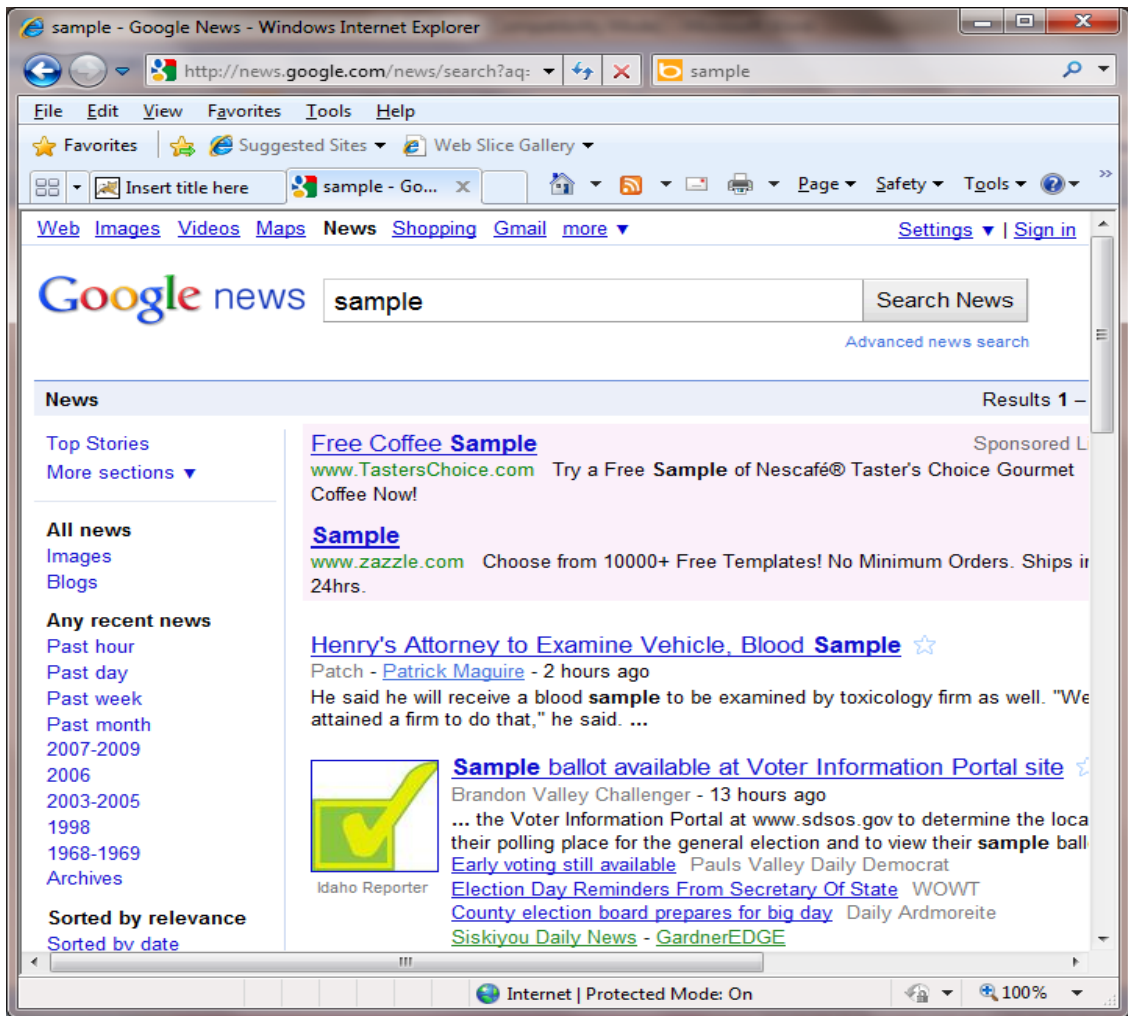

Fig 5.1 Original results of Google news search engine

Apply the "Own a Search Engine" technique, when user access a fake search engine and tried to search for a string "sample", the fake engine internally redirects the search request to the original news.google.com engine and retrieve the results and then modifies the branding of results to display own branding. Finally user will see the results of nesws.google.com search engine with the fake branding. Following figure explains the hacking of news.google.com search results.

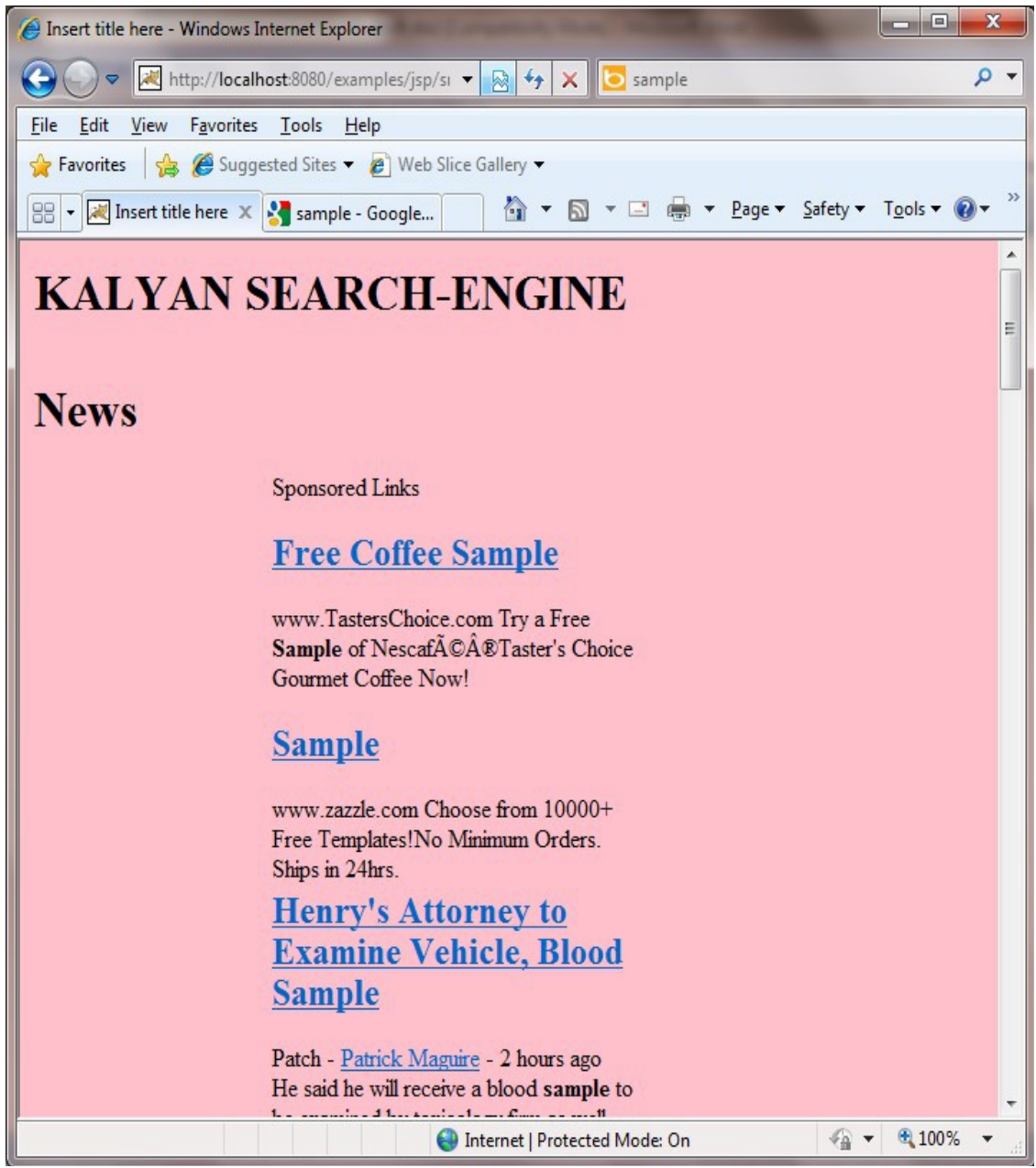

Fig 5.2 Fake engine internally uses Google-News search engine and displays the results with fake branding

Compare the search results displayed in fig 5.1and 5.2. It is observed that same results are shown in the both cases because fake engine internally uses Google-News search engine. From the above figures it is proved that Google-News search engine was open to any HTML page and any user can hack the results of Google-News and can show the news.google.com results with fake branding. Hence this paper is concluding that:

"Google-News was failed to protect its Search Engine",

But currently Google has strong protection for its search engine.

# 6. CASE STUDY-IV: OWN THE ENGINE "SEARCH.GOOGLE.COM"

In this section proposed technique "Own a Search Engine" is applied to hack the search results of search.google.com but Google has strong protection to save their search engine from this kind of attack. When we tried to apply the "Own a Search Engine" technique on seacrh.Google.com, an exception is thrown to the fake search engine with the HTTP 403 error code.

Following figure shows how the Google is protecting its search engine.

Fig 6.1 shows the exception thrown from the Google Search Engine to the fake engine

# 7. CONCLUSION

This paper conducts a test to the search engines Google, Yahoo and Bing with a proposed technique and proves the openness of the top 3 search engines. The proposed test technique 'Own a search engine' proves that Google strongly protects its search engine where Bing and Yahoo failed to protect their search engines from "Openness" flaw.

But Google also had loop holes in the past; applications like 'news.google.com' were suffered from Openness problem. As a conclusion, this paper proves a critical loop hole of top search engines called "Openness of a Search Engine" with a proposed technique. And this paper strongly appeals to Bing and Yahoo to fix the critical flaws of their Search Engines; else anyone can own the search results of their search engines.

**Authors**

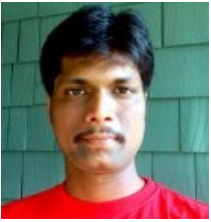

**Name: Katuru SM Kalyana Chakravarthy**

**Short Biography:**

"Katuru SM Kalyana Chakravarthy" is a Software Test Engineer, having 5+ years of experience in Software Testing and Quality Assurance Services. Currently he is working as Automation Engineer in Oracle India Private Ltd, Hyderabad, India and he has worked in AppLabs, India for five Years including one year of on-site experience. He is much interested in Research & Development, submitted white papers to the international conferences and presented a Poster in the conference ICISS 2009, Kolkata, India.